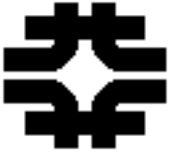

# Recycler Barrier RF Buckets

C. M. Bhat

Accelerator Division, Fermi National Accelerator Laboratory,

P. O. Box 500, Batavia, IL 60510, USA

(March 4, 2011)

(Submitted for Publications in a Book)

**Abstract:** The Recycler Ring at Fermilab uses a barrier rf systems for all of its rf manipulations. In this paper, I will give an overview of historical perspective on barrier rf system, the longitudinal beam dynamics issues, aspects of rf linearization to produce long flat bunches and methods used for emittance measurements of the beam in the RR barrier rf buckets. Current rf manipulation schemes used for antiproton beam stacking and longitudinal momentum mining of the RR beam for the Tevatron collider operation are explained along with their importance in spectacular success of the Tevatron luminosity performance.

"Work supported by Fermi Research Alliance, LLC under Contract No. DE-AC02-07CH11359 with the US Department of Energy."

## I. INTRODUCTION

Barrier bucket rf systems in synchrotrons were not invented by accident or serendipity but by their sheer necessity. At the very early stages of the Tevatron I Project [1], it was realized that the circumference difference of about 30 meters between the Fermilab antiproton Debuncher and the Accumulator Ring would result in an antiproton loss of about 7% for every transfer from the former to the latter ring. Therefore, it was essential to develop a technique to preserve a gap in the antiproton beam in the Debuncher with a minimum length equivalent to the circumference difference between the two synchrotrons. Furthermore, there was also a need to produce an isolated 2.5 MHz sinusoidal wave (harmonic number $h$ =4) to have a single rf bucket (isolated rf bucket) in the ring, with the rest of the buckets suppressed in order to extract a single antiproton bunch from the Accumulator ring for collider operation. These two requirements led to the initial development of barrier rf technology at Fermilab [2]. The concept of a "suppressed bucket" in synchrotrons has been addressed in the past [3]. Nevertheless, significant research was required on barrier rf (also called wide-band rf) systems to fulfil its initial needs and investigate other applications. As a result of this, many applications have been realized and now most of the synchrotrons at Fermilab are equipped with such rf systems [4]. The use of wide band rf systems as beam stabilizing tools has now become indispensible for high intensity beam operation at Fermilab. The Recycler Ring (RR) at Fermilab [5] uses only a barrier rf system in all of its beam manipulations, unlike any other storage ring in the world.

The RR is an 8.938 GeV/c$^2$ fixed energy antiproton storage synchrotron built using strontium ferrite permanent magnets. The Main Injector (MI) [6] and the RR share same underground tunnel with the RR attached to the ceiling of the tunnel. The lattice of the RR is essentially the same as that of the MI except at the straight section dedicated to the electron-cooling. The RR was intended originally as the main depository for i) antiprotons from the Accumulator Ring by periodic transfers, ii) recycled antiprotons from the Tevatron at the end of each proton-antiproton *store* (the recycled antiprotons expected have much larger 6D-emittance as compared to the one coming from the Accumulator Ring) and capable of providing cooled antiprotons for the Tevatron collider operation. The storage capacity of the RR was about $2.5\times10^{12}$ antiprotons [5] by design. It was highly essential to store and cool these two beams simultaneously in different parcels azimuthally around the ring and if required, merge them before transfer to the Tevatron for the subsequent proton-antiproton store. A state of the art rf technology with resonant rf systems was highly inadequate to fulfil all of these requirements. It was realized that the barrier rf technology was the most suitable solution for the RR. As the antiproton production rate from the pbar source was realized to be at least a factor of four higher than what was achieved during Tevatron Collider Run I, the original plan of recycling of the antiprotons from the Tevatron was dropped. Consequently, the RR was never used as an antiproton recycling storage ring.

## II. BARRIER RF WAVEFORMS

One can imagine a variety of barrier rf waveforms. A schematic view of typical barrier rf voltage excursions (red curves) for a synchrotron operating below $\gamma_T$ (like the RR) is shown in Figure 1(b)-(e) and compared with a standard sinusoidal rf voltage wave produced using a resonant rf system with harmonic number h=4 (Fig.1(a)). The barrier combinations shown in the first three examples are used for confining a parcel of beam in the synchrotron. Figure 1(e) is an example of a gap producing barrier combination, also referred to as "anti-buckets". A barrier rf wave of an arbitrary shape in a circular accelerator is the result of superposition of Fourier components of harmonics of the fundamental revolution frequency, $f_0 = 1/T_0$ where $T_0$ is the revolution period. Electronic generation of the barrier waveforms illustrated in Fig.1 is a straightforward low level rf (LLRF) electronics problem involving phase-lock loops, linear gates, etc. It has been shown [2] that an acceptable isolated sinusoidal waveform shown in Fig. 1(b) can be generated by taking only the first ten harmonic components of its Fourier expansion. The usable bandwidth of each component in the system should extend at least to about ten components above and below the required range of the Fourier components. For a better representation one needs a larger numbers of components. The RR barrier rf system uses >100 Fourier components to generate any waveform. The LLRF signals synthesized by this method are fed to a broadband rf cavity after an amplification using a broadband power amplifier. Below we give a general overview of dynamics of a charged particle in the presence of such a barrier rf wave in a synchrotron.

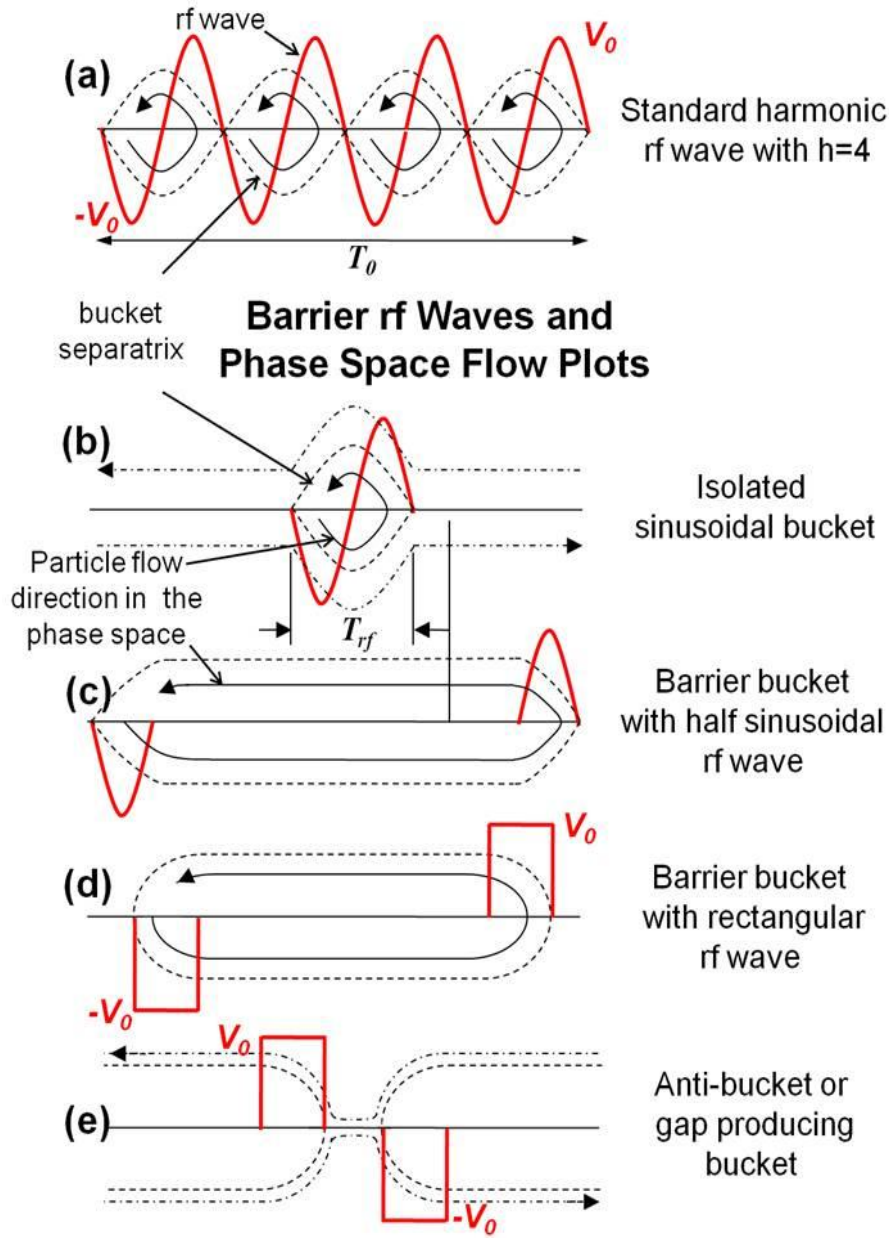

Fig. 1: A schematic view of a) a standard sinusoidal rf voltage wave produced using a resonant rf system with harmonic number h=4, b) an isolated sinusoidal wave c) two half sinusoidal waves with a gap, d) two rectangular waves with a gap and e) a gap preserving rf wave (anti-bucket). Red curves show rf wave forms in all of these cases. The dashed, solid and dashed-dot curves indicate bucket boundaries, particle flow directions in longitudinal phase space and flow directions of particles outside the buckets, respectively.

## III. LONGITUDINAL BEAM DYNAMICS OF CHARGED PARTICLES IN THE BARRIER RF BUCKETS – AN OVERVIEW

An off energy charged particle will continue to slip relative to a synchronous particle in a synchrotron in the region with zero rf voltage in a barrier bucket. It will lose or gain energy as soon as it encounters a rf barrier pulse and this will continue till there is enough kick from the barrier pulse to change its direction of slip. Thus, the barrier buckets sets the particles into oscillations. The motion of any particle with energy $\Delta E$ relative to a synchronous particle in a synchrotron is governed by [7, 8],

$$\frac{d\tau}{dt} = -\eta \frac{2\pi \Delta E}{T_0 \beta^2 E_0} \quad \text{and} \quad \frac{d(\Delta E)}{dt} = \frac{eV(\tau)}{T_0} \tag{1}$$

The quantities $E_0, e, \eta$ and $\beta$ are synchronous energy, electronic charge, phase slip factor and the ratio of the particle velocity to that of light, respectively. $-\tau$ is the time difference between the arrival of the particle and that of a synchronous particle at the centre of the rf bucket. $V(\tau)$ is the amplitude of the rf voltage waveform. Using the above equations, one can get the half bucket height $\Delta E_b$, given by,

$$\Delta E_b = \sqrt{\frac{2\beta^2 E_0}{|\eta|} \frac{\left|\int_{T_2/2}^{T_2/2+T_1} eV(\tau)d\tau\right|}{T_0}} \tag{2}$$

Since the bucket height depends upon $\int eV(\tau)d\tau$ the exact shape of the wave form is not very critical. The quantities $T_1$ and $T_2$ denote barrier pulse width and gap between rf pulses, respectively, as indicated in Fig. 2.

In the case of the RR, one mostly deals with rectangular barrier buckets of the type

$$V(\tau) = \begin{vmatrix} -V_0 & \text{for } -T_1 - T_2/2 \leq \tau < -T_2/2 \\ 0 & \text{for } -T_2/2 \leq \tau < T_2/2 \\ V_0 & \text{for } T_2/2 \leq \tau < T_1 + T_2/2 \end{vmatrix} \tag{3}.$$

The longitudinal beam dynamics in such a barrier bucket is less understood. Progress is made very recently [8-10] which were mainly driven by RR requirements. The advantage of a rectangular waveform is that for a given maximum rf voltage $V_0$, the available bucket area will be a maximized. A schematic view of the beam phase space distribution in a rectangular barrier rf bucket along with the definition of various parameters relevant to this document is also shown in Fig. 2.

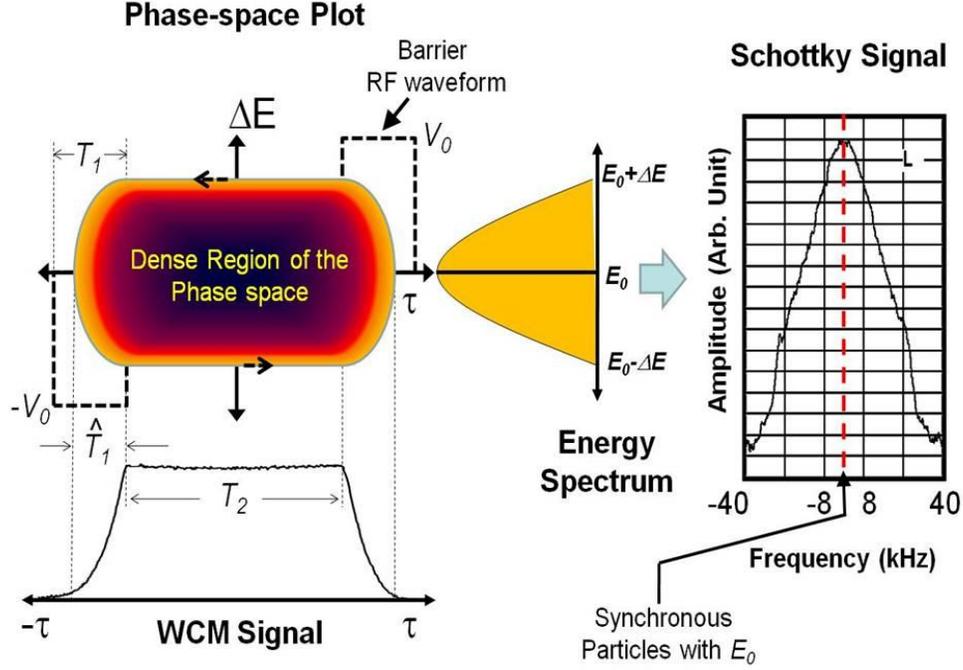

Fig. 2: A schematic of ($\Delta E, \tau$) –phase space distribution of particles in a rectangular barrier rf bucket with its line-charge distribution (wall current monitor-WCM signal) and energy distribution. The typical Schottky spectrum of such a beam bunch is also shown.

It has been shown [8-10] that the energy offset $\Delta \hat{E}$ of a particle is related to its depth of penetration $\hat{T_1}$ into a rectangular barrier by,

$$\Delta \hat{E} = \sqrt{\frac{2\beta^2 E_0}{|\eta|} \frac{eV_0 \hat{T_1}}{T_0}} \tag{4}$$

and the total longitudinal emittance $\varepsilon_l$ of the beam particles with $\Delta E \leq \Delta \hat{E}$ is,

$$\varepsilon_l = 2T_2 \Delta \hat{E} + \frac{8\pi|\eta|}{3\omega_o \beta^2 E_o eV_o} \Delta \hat{E}^3 \tag{5}$$

where $\omega_0 = 2\pi/T_0$. The synchrotron oscillation period of the particle is,

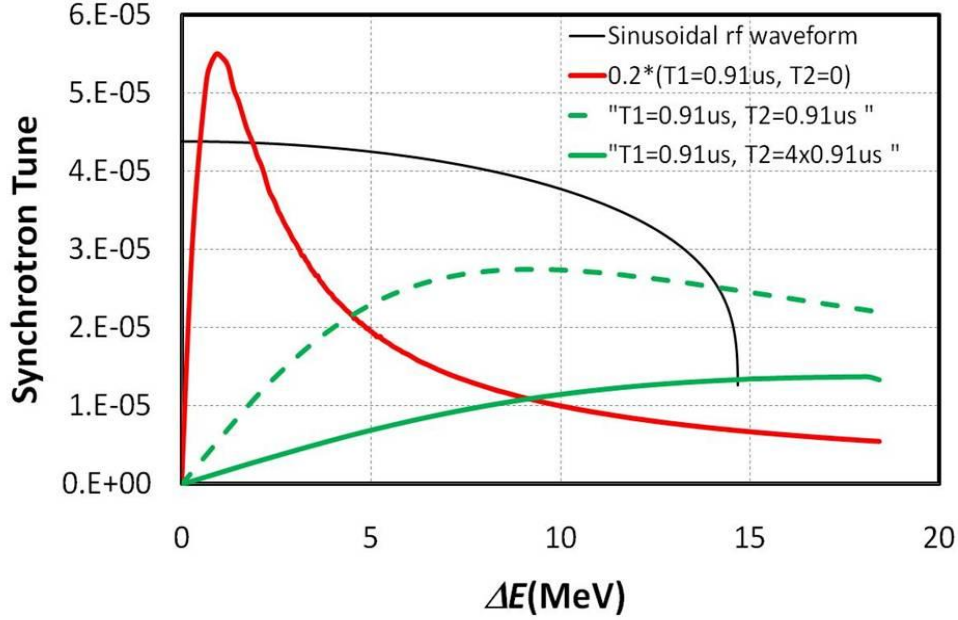

Fig. 3: RR synchrotron tune versus energy offset of the particle in rectangular barrier buckets for three scenarios of $T_2$ and their comparison with a sinusoidal rf bucket of length $2T_1$ (black solid line).

$$T_s = \frac{2T_2}{|\eta|}\frac{\beta^2 E_o}{\left|\Delta \hat{E}\right|} + \frac{4\left|\Delta \hat{E}\right|T_0}{eV_o}; \text{synchrotron frequency } f_s = \frac{1}{T_s} \qquad (6).$$

The Eq. (6) has two components; one for the region with $V(\tau)=0$ and the second for $V(\tau)\neq 0$. Figure 3 displays a comparison between the calculated synchrotron tune ($Q_s = T_0 f_s$) spectrum for particles in a sinusoidal rf bucket in RR (black solid line) with a period of $2T_1$ and that in a barrier bucket with $T_1 = 0.91$ μs and $T_2 = 0.0$ (red line), 0.91 (dashed green line) and 3.64 μs (green line). In all these cases $V_0 = 2$ kV. It is interesting to note that the synchrotron frequency spectrum for the particles in a rectangular barrier bucket is significantly different from that for a sinusoidal rf bucket. In the case of the barrier buckets, the peak energy offset at which the synchrotron period reaches minimum is $\Delta \hat{E}|_{T_s (Min)} = \sqrt{T_2/4T_1}\Delta E_b$. Consequently, if $T_2 \leq 4T_1$ then the $\frac{df_s}{d(\Delta E)} = 0$ lies well inside the bucket boundary. The particles in the vicinity of $Q_{sMax}$ loose Landau damping [10] and beam may become susceptible to longitudinal collective instability and/or longitudinal instability driven by external noise sources.

The longitudinal beam dynamics in barrier buckets explained above is well modeled in a computer code ESME [11]. All rf manipulation schemes were simulated prior to their use in the RR.

## IV. RR RF SYSTEM

RR rf specification calls for the beam to be contained in barrier buckets made of rf pulses of different wave shapes, lengths and $V_0$ in one or more parcels azimuthally distributed around the ring. Further these parcels should be capable of moving relative to one another and should be capable of expansion and compression independent of one another. The rf system is designed to meet all needs required by the RR.

The RR rf system is comprised of a LLRF and a high level rf (HLRF) system. The RR LLRF [12] is a very advanced and versatile control system consisting of a Super Harvard Architecture Computer (SHARC) digital signal processors (DSPs). In fact, for Run II the LLRF systems in most of the accelerators at Fermilab like the Booster ring, MI, Tevatron and RR have been upgraded to adopt similar architecture. The RR LLRF uses ADSP-2106x SHARC high performance floating point DSP. The SHARC derives much of its performance like, generating data address, hardware loop control and all multifunction instructions in parallel. All instructions are executed in one clock cycle. The key feature of this system is at 66 MHz clock rate the DSP core peaks at 198MFLOPS. The DSP (8-bit) along with its software controls the frequency and phase registers of three direct digital synthesizer (DDS) channels. These DDS modules provide nine rf clock inputs to a new SHARC based module called the Recycler Bucket Generator (RBG). The RBG contains eight arbitrary waveform generator channels (often called ARB). The DSP controlled ARBs are summed to form the LLRF output that drives Recycler HLRF system. Thus, each arbitrary waveform generator has an independent table, phase, and amplitude, all of which can be changed in real time to perform a specific type of rf manipulation.

The RBG faces many new challenges as compared with any previous LLRF controls used at Fermilab. Each barrier pulse should be capable of turning on and off independent of other pulses. To meet these requirements each waveform channel has its own clock controlled by DDS. Each DDS is programmed to the same frequency, while the phase is programmed to advance or retard the barrier pulse. Any change to the shape of the barrier pulse may be made at a 720 Hz rate. The LLRF commands for barrier rf manipulations are passed from the CPU over a VXI bus to RBG. The RBG then smoothly controls any changes to bucket parameters. Table 1 lists available *cogging*[1] rates for any rf pulses in the RR. All of these cogging rates are used in the Recycler operation. For example, the "slow" rate is used when the beam emittance need to be preserved at the level of a few percent, irrespective of the spread of the synchrotron oscillation period of the particles in it. These features were more than sufficient for the RR operation.

Table 1: RR LLRF cog rate for the barrier pulses [12]

| Cogging Param | Medium | Slow | Fast |
|---|---|---|---|
| Max Rate (Bkt*/sec) | 39.98 | 4.86 | 8400 |
| Max Derivative (Hz/sec) | 12.5 | 1.94 | 3024000 |
| Min. Cogging time (sec) | 6.9 | 5.08 | 0.01389 |
| Max. Bkts cogged in one minute | 137.996 | 12.355 | ~44000 |

*Bkt=18.935 nsec, 53MHz bucket length

In the current RR LLRF architecture ARB5 (out of ARBn, n= 0,1,...7) is used as RR reference marker for synchronous beam transfers to and from the RR and for cogging the rest of the ARBs. Each of the ARBs are assigned with a predefined rf wave form which can be turned on and off in 128 steps of equal amplitudes at variable rates. There are 16 different waveforms in use. A rectangular positive barrier pulse with a width of 0.908 μs, a similar barrier pulse with negative excursion or a 2.5 MHz sinusoidal rf wave with a period of four, etc. are examples of predefined rf waves. The maximum width of a waveform is limited to 256 Bkt (4.84 μs).

---

[1] This is a quasi-adiabatic process that involves azimuthal displacement of a bunch in a synchrotron. Cogging is carried out by a continuous change in phase and/or frequency of an rf wave relative to synchronous particles.

The RR LLRF architecture explained above is quite general. Its full potential was realized by developing new sets of commands specific to RR operation [13]. For example, "UpdateArbWaveform" is a command which can be used to assign any of the eight ARBs to a waveform. A command like "GrowCoolBucket is used to expand a cold bucket by a certain amount; here a cold bucket is defined in a specific way. Currently, there are about 75 such commands in use, each for a specific purpose. Each rf manipulation scheme, like antiproton beam stacking, is treated as a group of such commands put together in a module.

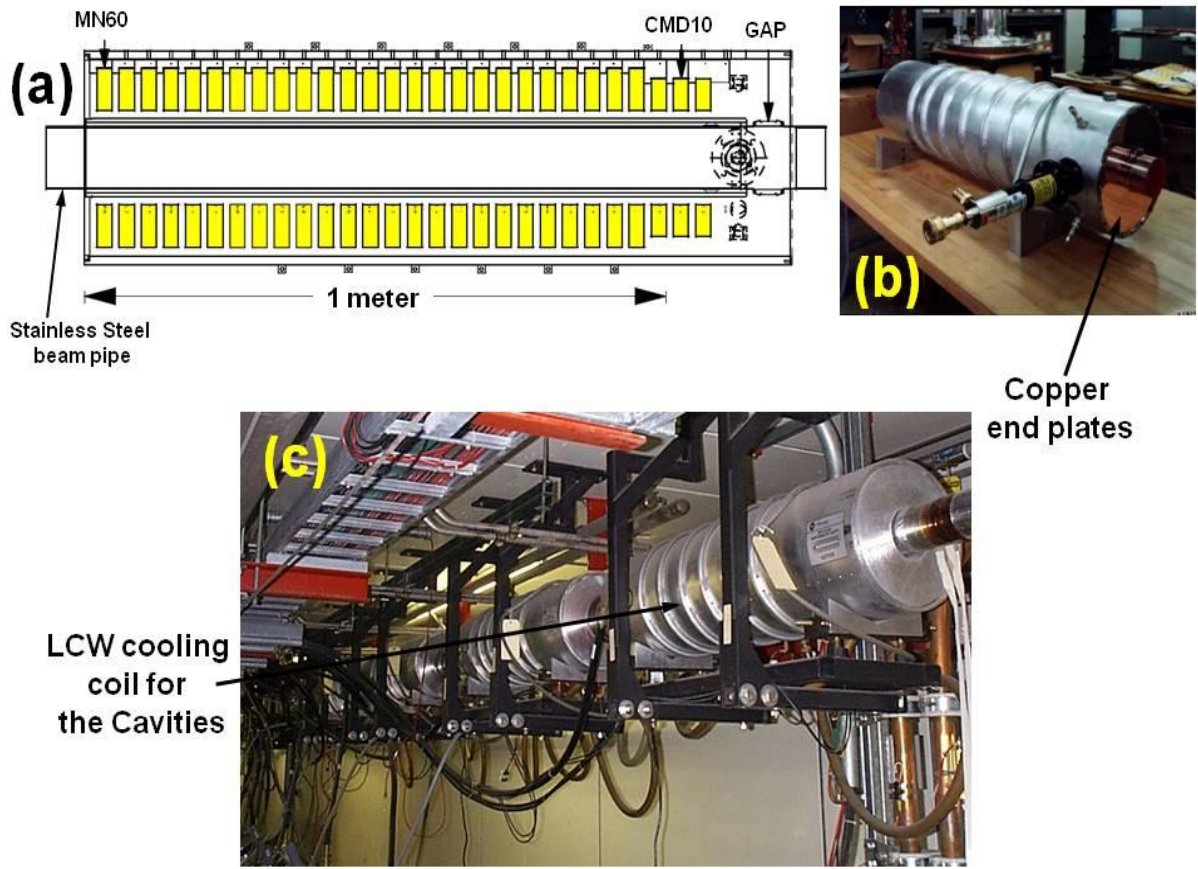

Fig. 4: RR barrier rf cavity [14]: a) a schematic of the cavity showing all of the internal parts of a single cavity, b) cavity on a test bench with 60 Ω rf load and a copper end plate connecting to the beam pipe and c) after installation of the four cavities in the RR.

The RR HLRF [14] consists of four 50 Ω cavities, each driven individually by a broadband solid-state amplifier (Amplifier Research Model 3500A100). Each amplifier, operating in push-pull mode, can supply a minimum of 3500 Watts output power over the frequency range of 10 kHz to 100 MHz. The amplifiers are connected to the cavities in the tunnel by 20 m of 7/8" diameter coaxial line.

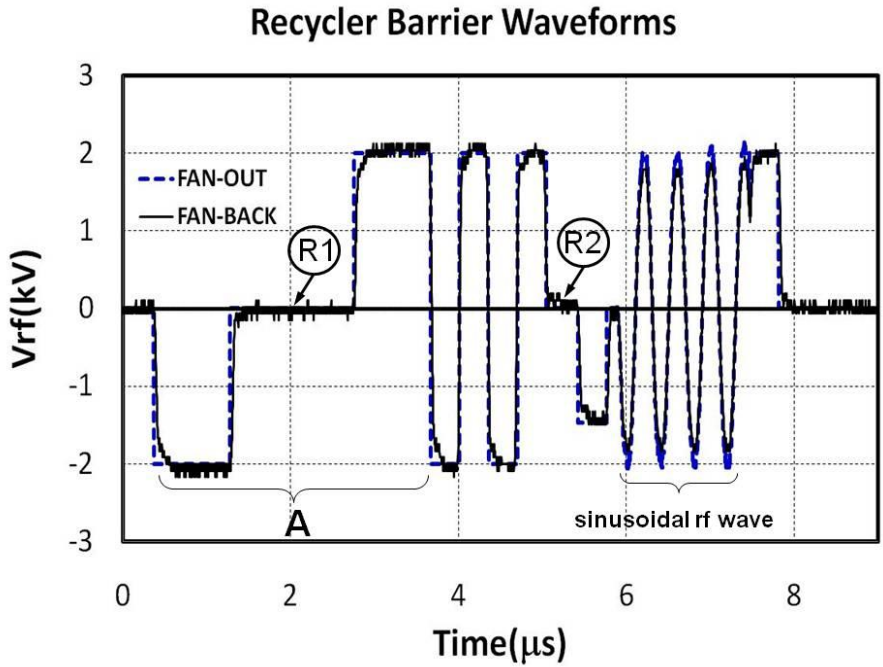

Fig. 5: A typical example of a complicated LLRF rf fan-out (dashed blue curve) and HLRF fan-back from the cavity pickups (black curve). The data shown is for about 80% of the ring.

Figure 4(a) shows an assembly drawing of a RR barrier rf cavity. Each cavity consists of twenty-five Mn-Zn ferrite (Ceramic Magnetics MN60) rings of dimension 11.5" OD × 6" ID × 1" thick with a spacing of 0.5" between each ring. Each ring is supported by Kapton spacer blocks. The Mn-Zn ferrite becomes very lossy above about 1 MHz. To extend the high impedance range above 10 MHz the assembly is augmented with three additional rings of NiZn ferrite (CMD10) as shown in the figure. The whole assembly is enclosed between concentric aluminium cylinders encircling a 4" diameter stainless steel beam pipe with a 1" ceramic gap. This gap is electrically connected to the cavity with beryllium-copper finger stock which couples the developed electric field to the beam. To complete the electrical connection between the beam pipe and the cavity outer aluminium shell, a thin copper end plate is used (see Fig. 4(b)). The cavity is cooled using low conductivity water (LCW) flow through a coil wound around the outer shell whereas the ferrite core is air cooled. The impedance of the cavity is approximately 50 Ω over a frequency range of 100 kHz to 20 MHz when shunted with a 60 Ω resistor. The designed peak voltage per cavity is 500 V. Figure 4 (c) shows a picture of four barrier rf cavities in the RR.

Figure 5 shows an example of the input LLRF fan-out signal and the HLRF fan-back sum signal measured for the four RR cavities. The combined maximum available peak gap voltage is about 2 kV. A small discrepancy between the fan-out and fan-back at the rising (falling) edge of a positive (negative) barrier pulse was inherent to this system; this did not limit the performance of the RR.

**V. RF LINEARIZATION**

It is extremely important to have a flat longitudinal line charge distribution in the region with $V(\tau) = 0$ for the rf barrier bucket (Fig. 1(c), (d) or 2). But the beam profiles in the RR barrier buckets with $T_2 \neq 0$ (in absence of any compensation) showed clear unevenness. This lead to unequal intensity and emittance for bunches sent to the Tevatron. The main causes for unevenness in the line charge distribution are i) nonzero harmonic contents of the revolution frequency between the positive and the negative barrier pulses defining the bucket, ii) polar asymmetry in the fan-back signals of barrier pulses, iii) potential well distortion and beam loading and iv) rf imperfections. Each of these effects was observed in the RR as beam cooling was improved.

A non-zero harmonic component of the revolution frequency was seen for the first time [15, 16] during beam measurements on an isolated rectangular barrier bucket similar to one indicated by "A" in Fig. 5. Ideally, the fan-back signals in regions "R1" and "R2" in Fig. 5 should be zero irrespective of the exact shapes of the neighbouring barrier pulses and the presence of other barrier buckets. Measurements of the frequency response on the fan-back signal of such a waveform showed that there was a considerable amount of non-linearity in magnitude and phase, most of which came from the solid-state rf amplifiers. A consequence of this was head-tail asymmetry in the beam profile even at very low beam intensity as shown in Fig 6(a). A linearizing circuit made of high pass, band pass and low pass filters was added between the LLRF output and the cavity amplifiers [17]. The current system linearizes the frequency in the region from 90 kHz to 1 MHz (up to ten harmonics) with flatness in amplitude better than 0.26 dB and phase $1.8^0$. Figure 6(c) shows the wall current monitor (WCM) signal after the corrections.

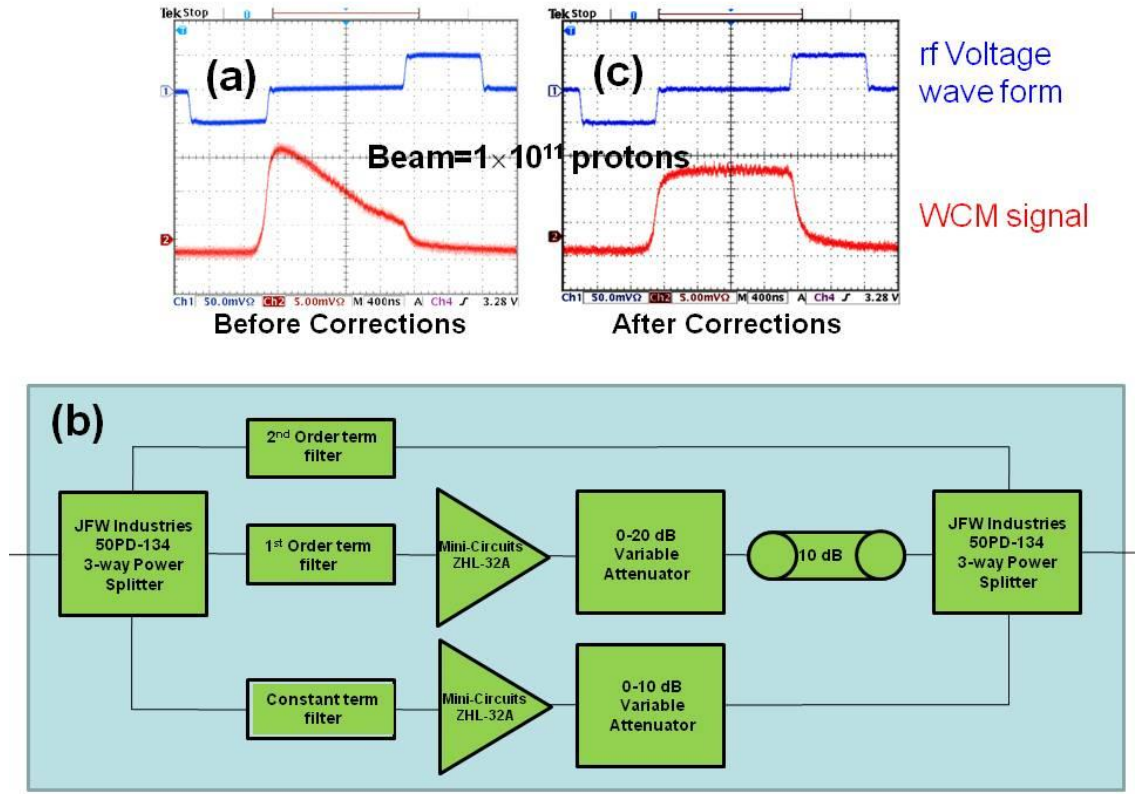

Fig. 6: RR barrier bucket (blue) and beam profile (red) a) before correction b) block diagram of the linearizing circuit c) after the corrections [17]. The horizontal axis is time (400 nsec/div)

The origin of the polar asymmetry for barrier rf pulses was identified to be non-symmetric saturation curves of the power amplifiers; these saturation curves were unique to each amplifier. Consequently, the measured difference between the absolute values of fan-out signal for a positive and a negative rf pulse was about 7% in the case of a rectangular rf bucket shown in Fig. 5. Furthermore, this difference is a function of the position of barrier pulse relative to the other in a bucket and the presence of additional barrier buckets in the ring. A temporary solution was to impose $\int [V_+(\tau) - V_-(\tau)] d\tau = 0$ on the LLRF fan-out inputs until a more robust, FPGA based, correction (explained later) was developed.

As the RR beam got colder the observed distortion of the beam profile between barriers found to resemble the inverse of the potential-well $\int V(\tau)d\tau$ as shown in Fig. 7(a). A theoretical study of this unevenness arising from potential-well distortion was carried out using Haissinski equation [18, 19] which describes the observed beam profile as a function of time according to,

$$\rho(\tau) = \rho_0 \exp\left[-\frac{|e|\beta^2 E_0}{|\eta|T_0\sigma_E^2}\int_0^\tau V_{eff}(t)dt\right] \tag{7}$$

Here $\rho_0$, $\sigma_E$ and $V_{eff}(t)$ are the ideal profile of the beam, measured root mean square energy spread and fan-back voltage, respectively. The net distortion of the beam current distribution, $(\rho(\tau)-\rho_0)$ can be obtained by expanding Eq. (7) as,

$$\rho(\tau) - \rho_0 = \rho_0 \frac{|e|\beta^2 E_0}{|\eta|T_0\sigma_E^2}\int_0^\tau V_{eff}(t)dt \tag{8}$$

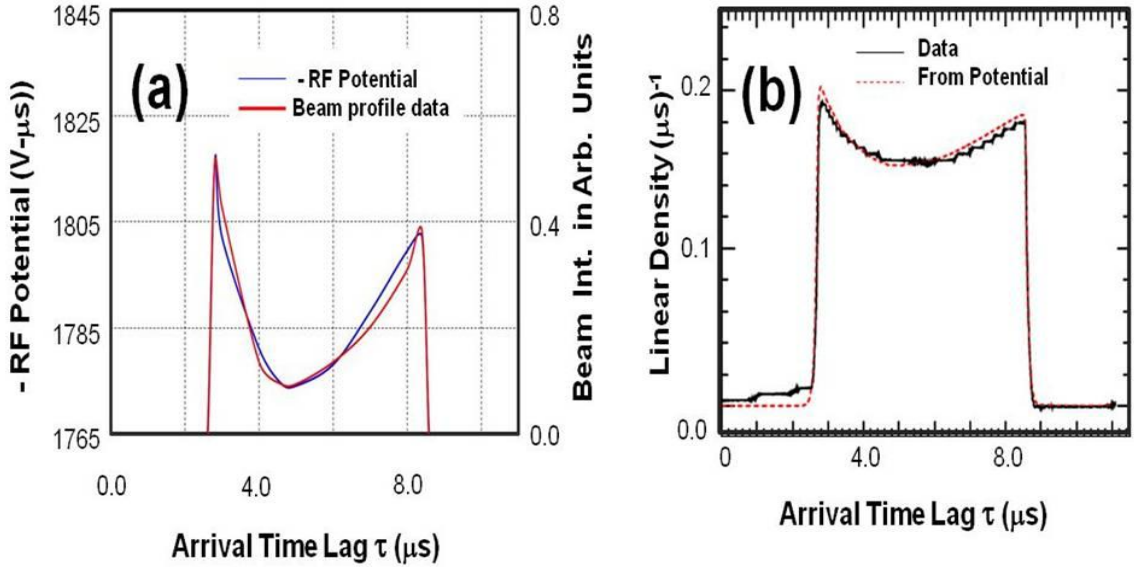

Fig. 7: a) Line charge distribution of cold antiproton beam ($0.51\times10^{12}$ antiprotons) in a rectangular barrier bucket overlapped on the inverse of the rf potential, b) comparison between computed beam profile using Haissinski equation and measurements [19] for the same beam.

This equation clearly explains one of the important features of experimental observations *viz.*, inverse dependence of distortion on $\sigma_E^2$. Fig. 7(b) shows a comparison between predicted (red dashed curve) and measured (dark continuous curve) beam intensity profiles for $0.51 \times 10^{12}$ antiprotons in a rectangular barrier bucket with $T_2 = 5.8$ μs. Almost all of the observed unevenness in this case was due to rf imperfections with a very small contribution from the beam loading. The beam loading effect showed up at around $1 \times 10^{12}$ antiprotons under similar conditions. Further studies showed that the magnitude of the rf imperfection also depends upon $T_2$, the presence of other buckets, their size and dynamics of barrier pulses.

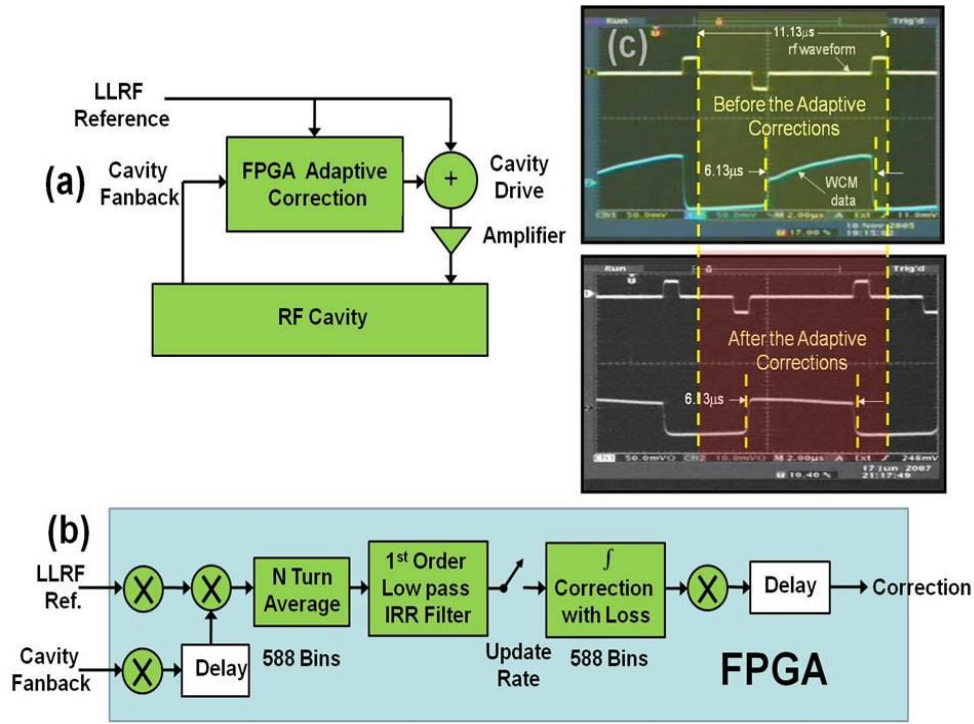

Fig. 8: (a) Block diagrams [20] for a) the adaptive correction system implemented in the RR LLRF, b) correction algorithm inside the FPGA firmware, and c) rf fan-back signals and WCM beam profile before (top) and after (bottom) corrections on a $3.1 \times 10^{12}$ anit-proton captured in a rectangular barrier bucket of $T_2 = 6.13$ μs. The extent of the RR indicating one revolution period (11.13 μs), is shown by the light shaded area.

To address these problems once and for all, a FPGA-based adaptive correction system was developed [20]. The correction system uses an average error signal obtained by taking the difference between the fan-out and fan-back signals over 128 turns. This signal contains the required information on the observed rf imperfection and beam loading. The final corrections generated using a FPGA firmware is then summed back into the LLRF reference signal by an external summing amplifier board. Figs 8(a) and (b) show block diagrams representing the adaptive correction system implemented in the RR. Fig. 8(c) shows a typical case of beam profiles before and after the FPGA corrections. This correction system has become an integral part of the RR operation and was found to be very effective for all rf manipulations carried out in the RR, even on the cold antiproton intensities $>5\times 10^{12}$.

## VI. MEASUREMENTS OF LONGITUDINAL AND TRANSVERSE EMITTANCE

A precision measurement of the emittance of the beam in a particle accelerator is a key factor in understanding the beam dynamics. In the past, both destructive as well as non-destructive techniques have been employed to measure the emittance. The RR uses only non-destructive or semi non-destructive techniques for emittance measurements.

The Recycler uses two independent emittance measurement techniques in the longitudinal and transverse planes. One of the two methods in the RR is based on the Schottky signal measurements. The use of Schottky signals for 6D- emittance measurements is an old and elegant non-destructive method for storage rings [21]. Originally, this technique was used for a coasting beam. However, for a bunched beam it has been shown that [22] if $T_M f_s \ll 1$ where $T_M$ is duration of measurement (characteristic time of the band-pass filter), the spectrum measured with a longitudinal Schottky detector of proper operating frequency represents an instantaneous "snap-shot" of the momentum distribution of beam particles.

A detailed account of various longitudinal emittance measurement methods adopted in the RR prior to 2003 is described in Ref. [23]. The current technique adopted in the RR is capable of measuring the longitudinal emittances for a segmented beam captured in two barrier buckets by gating the Schottky signals [24]. The energy spread of the beam $\Delta \hat{E}$ (rms and/or 95% value) is measured using the longitudinal Schottky spectrum of the beam. The barrier pulse gap $T_2$ is read from the LLRF fan-out signals. The longitudinal emittance is obtained using Eq. 5. The accuracy of the measured longitudinal emittance by this method on a bunch with $T_2 \geq 4T_1$ is found to be better than ±15%. More accurate offline methods based on beam tomography using Schottky data and/or WCM data [23, 25] have been developed. The accuracy from these methods is better than ±10% even for shorter bunches.

Flying wires and transverse Schottky detectors are used to measure the transverse emittance ($\varepsilon_T$). The flying wires measurements have accuracy $\approx$ ±15% but each fly gives rise to a few percent emittance growth on the antiproton beam (on a very cold beam the emittance dilution by flying a wire can be significantly large). As such, it can not be used very often. On the other hand, the Schottky method is completely non-destructive, but, the measurement accuracy depends on the length of the barrier bucket. The error could be better than 5% on coasting beam or on a long bunch. The errors are as large as 100% on short bunches. It is important to note that this method relies on the measurement of the power spectrum from the Schottky noise of the beam in barrier buckets which could introduce very large uncertainties for very small values of $T_2$ (due to contributions from coherent signals).

During normal RR operation, the Schottky method is used to measure relative values of the emittances. More accurate values of the emittances are obtained from the flying wire measurements for transverse planes and a method based on WCM data [25, 26] in the longitudinal planes.

## VII. ANTIPROTON STACKING

The antiprotons from the Accumulator Ring are transferred to the RR via the MI in the form of four 2.5 MHz bunches for every transfer. There is a small frequency mismatch between the Accumulator Ring and the MI and, energy mismatch between the Accumulator Ring and the RR. As a result of this, the MI is made to play the role of an intermediate injector synchrotron rather than a transfer line, in contrast to the original design [5]. To eliminate any significant emittance dilution while transferring the beam between these three rings a frequency shift[2] of about 4550 Hz (of 53 MHz rf system of the MI) [27] between the Accumulator Ring and the MI was introduced just before the injection. The antiprotons are decelerated in the MI by about 40

---

[2] A frequency shift is introduced momentarily in MI only to synchronize Accumulator Ring and the MI. Just before the antiprotons transfer the frequency shift is removed.

MeV/c before it is transferred to the RR. Nevertheless, an emittance dilution at the level of a few percent is inevitable between the Accumulator Ring and the RR. As a consequence of this, there is a correlation between the overall longitudinal emittance dilution and Accumulator Ring to RR stacking efficiency $\mathcal{E}_{AR \to RR}$, given by,

$$\mathcal{E}_{AR \to RR} = \frac{(\text{RR Beam})_{\text{Final}} - (\text{RR Beam})_{\text{Initial}}}{\text{Antiprotons from the Accumulator Ring}} \qquad (9).$$

The correlation arises due to the limited area of the barrier bucket in the RR available for stacking. If the beam particles fall out of the barrier bucket in any stages of the beam stacking, they will become DC and almost 30% these antiprotons will be lost during each transfer. Therefore, the total Accumulator to RR antiproton transfer efficiency is looked upon as a product of efficiencies between i) Accumulator Ring to MI and ii) MI to RR. Here we emphasis on the latter component.

Over the years, a number of quasi-adiabatic rf stacking techniques have been developed [28] with an emphasis on maximizing the overall stacking efficiency. The technique currently being used is explained below.

Figure 9 shows various steps of rf manipulation used during antiproton stacking. Prior to the beam transfer an anti-bucket is grown adiabatically to keep the injection region out of any antiprotons with $17 \text{ MeV} \leq |\Delta E| \leq 34 \text{ MeV}$; the particles with $|\Delta E| < 17 \text{ MeV}$ are confined to the old stack bound by 2 kV barrier pulses. Nearly a second before the beam transfer to the RR a set of four 2.5 MHz rf buckets of length $\approx 1.59 \text{ μs}$ is superimposed inside the anti-bucket, as indicated by "A" in Fig. 9(a). Following the beam injection, the anti-bucket is replaced by a standard capture barrier bucket ("B" in Fig. 9(b)). Next, the newly arrived beam is debunched by removing the 2.5 MHz rf buckets and the $|\hat{\Delta E}|$ is matched to that of the old stack before merging them using a *morph*-merging technique [28]. Figure 9(c) shows an intermediate step of this combing process. The beam after merging is shown in Fig. 9(d) with an anti-bucket grown in the injection region for the next transfer. This procedure is repeated as many times as needed to complete a set of beam transfers.

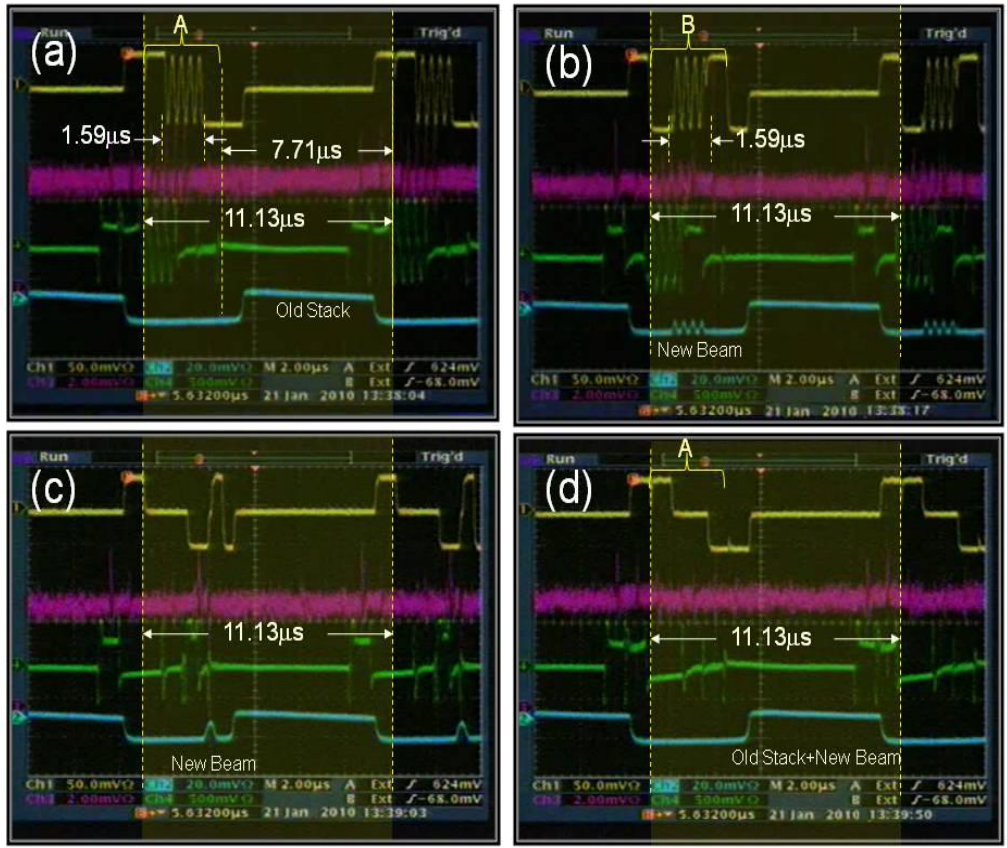

Fig. 9: Scope pictures for antiproton stacking in the RR. Each picture has four traces. They are HLRF-fanback (top, yellow), error signal from LLRF drive-fanback (2$^{nd}$ from top, purple), correction to the LLRF drive (green) and WCM data (bottom, light blue). The beam injection region with anti-bucket ("A" in (a) and (d)) and capture bucket ("B" in (b)) are also shown. For other details see the text.

Antiproton stack size in the RR is built up to a required size over several beam transfers by extracting almost all of the newly accumulated antiprotons from the Accumulator Ring. A set of transfers (a set consisting of 2-3 transfers each) will be carried out once for every 30 min and will have an average of about $0.25 \times 10^{12}$ antiprotons with large emittance – transversely about 6-7 $\pi$ mm-mr and longitudinally about 25 eVs. Between two sets transfers, the antiproton beam in the barrier bucket made of old stack and newly arrived antiprotons is cooled using Stochastic cooling systems (rarely with electron-cooling [29]; electron-cooling is generally used just before beam transfer to the Tevatron). Simultaneously, new antiprotons are accumulated in the Accumulator Ring for the next set of transfers. Typically, 15-20 sets of transfers is required to reach an optimum stack size[3] of $3.5\times10^{12}$ antiprotons in the RR.

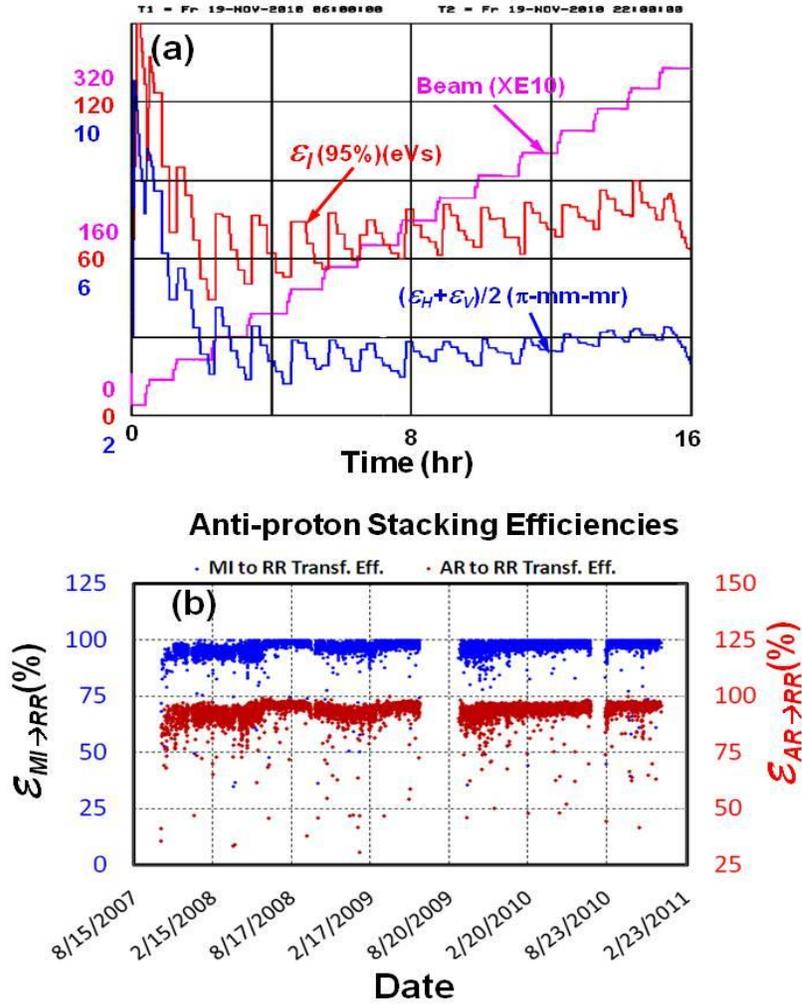

Fig. 10: (a) A typical example of RR antiproton stack build up along with measured longitudinal and transverse emittances. (b) Measured $\mathcal{E}_{MI\rightarrow RR}$ (blue data points) and $\mathcal{E}_{AR\rightarrow RR}$ for the last 40 months. The stacking technique used during this period is explained in the text. The intermediate breaks in "b" represent long shutdowns for the collider complex.

The emittance of the beam in the RR during stacking is shown in Fig. 10(a). The initial emittance of the leftover antiproton beam from the previous Tevatron load was quite large for the example shown here. Each step in the beam intensity represents a new set of transfers. The measured step increment in the beam emittance at the end of each set of transfer is approximately the same as that measured in the MI. This clearly implies that the rf manipulation used for stacking is almost adiabatic. The maximum stack size of the antiprotons obtained in the RR is $\approx$ $5.4\times10^{12}$ with $\varepsilon_l$ (95%)~70 eVs, average $\varepsilon_T$ (95%) ~ 3.4$\pi$-mm-mr.

---

[3] An optimum RR stack size is determined by several factors like the Tevatron luminosity life time, integrated luminosity delivered to collider detectors over a period of a week and also the rate at which the antiprotons are accumulated in the Accumulator Ring.

Figure 10(b) shows the measured $\varepsilon_{MI\to RR}$ and $\varepsilon_{AR\to RR}$ for the past three and half years. The $\varepsilon_{MI\to RR}$ is found to be nearly 100% with an average value of 95% over this period. The average of $\varepsilon_{AR\to RR}$ was about 92% for the same period. Around the middle of 2008, a change was introduced to the anit-bucket configuration in the RR which resulted in improved $\varepsilon_{MI\to RR}$ values. The data shows that these two efficiencies are closely correlated. The few percent difference between $\varepsilon_{MI\to RR}$ and $\varepsilon_{AR\to RR}$ is mainly dominated by the Accumulator Ring to MI beam transfer efficiency as explained earlier.

At this point it is quite important to make a remark on the effect of the MI on the RR beam. Early on an emittance growth in the stacked RR beam was is detected [15] which was arising from the leaked magnetic field from the MI acceleration cycles. Consequently, 1) a special task was undertaken to shield the RR beam pipe using mu-metal all around the ring and 2) a dynamic orbit compensation technique was developed (using RR correctors) and implemented in operation to negate the effect of MI cycles on the RR beam [30]. They helped to improve the RR performance significantly. As the electron cooling became operational transverse instability was observed in the antiproton stacks at very high phase space densities. To control the observed instability a damper system is installed with an initial bandwidth of 30 MHz and upgraded to a bandwidth of 70 MHz [31]. The maximum stack size of the antiprotons in RR with the current transverse damper is $\approx 5.4\times10^{12}$ with $\varepsilon_l$ (95%)~70 eVs, average $\varepsilon_T$ (95%) ~ 3.4$\pi$-mm-mr.

**VIII. ANTIPROTONS FOR THE COLLIDER OPERATION: LONGITUDINAL MOMENTUM MINING**

The steps involved in filling up the Tevatron with antiprotons are i) extract four 2.5 MHz bunches of antiprotons at a time from the RR (or from the Accumulator Ring or a mixture [32]) and send them to the MI and ii) accelerate them from 8 GeV to 150 GeV before transferring to the Tevatron. This step is repeated nine times to fill the Tevatron with thirty six antiproton bunches. The main requirements for Run II are 1) an efficient and robust way to extract low emittance antiproton bunches from the dense region of the phase space of the cold beam stack and leave behind the particles with very large $\pm \Delta E$ for use after cooling (called "antiproton economy") and 2) all 36 antiproton bunches in the Tevatron should have the same emittance and same number of particles per bunch, i.e., there should be no bunch to bunch variation in the proton-antiproton collider luminosity. To meet these requirements a new mining technique called longitudinal momentum mining (LMM) was developed [33]. LLM was crucial to the success of the RR even before the e-cooling was commissioned early July 2005. From middle of 2004 till October 2005 Tevatron stores were carried out in *mixed mode* [32] (a few bunches from the Accumulator Ring and rest from the RR; RR used LMM). From October 2005 till now the antiprotons came from LLM in the RR. Prior to the implementation of LMM, we simply sliced the cold antiproton distribution nine times along the time axis using another set of barrier bucket. This led to enormous emittance growth in longitudinal phase space. The overall emittance growth of the antiproton stack was on the order of 300% from the $1^{st}$ to $9^{th}$ transfers. In addition, later transfers suffered from lower bunch intensities and larger longitudinal emittance leading to a large antiproton loss in the RR as well as in the Main Injector during acceleration and bunch coalescing. Similar effects were seen for the antiprotons directly extracted from the Accumulator Ring for the Tevatron stores (for example, we had this problem throughout the Run I of the collider operation). As a result of these problems the maximum proton-antiproton luminosity delivered to the HEP program was severely limited.

The general principle of longitudinal momentum mining is illustrated in Fig. 11. The rf waveform along with the beam phase space boundary (dashed lines in left figures) and the corresponding potential well containing beam particles are shown for various stages of the mining processes. The objective of longitudinal momentum mining is to isolate particles closer to $E_0$ (i.e., dense region near the bottom of the potential well, Fig. 11(a)) from the rest without any emittance growth. This is accomplished by adiabatically inserting a set of mining buckets (the illustration in figure is for three parcels). The particles which cannot be bound by these "mini" barriers are still bound by the larger barriers shown in Fig.11(b) and are executing synchrotron oscillations at a relatively higher rate than the ones captured in the mini-buckets. Finally, the un-captured particles are isolated in another rf bucket (high momentum bucket) as shown in Figs. 11(c) and 11(d). Thus, are the particles in the dense region of the ($\Delta E, \tau$)-phase space are mined while leaving the rest.

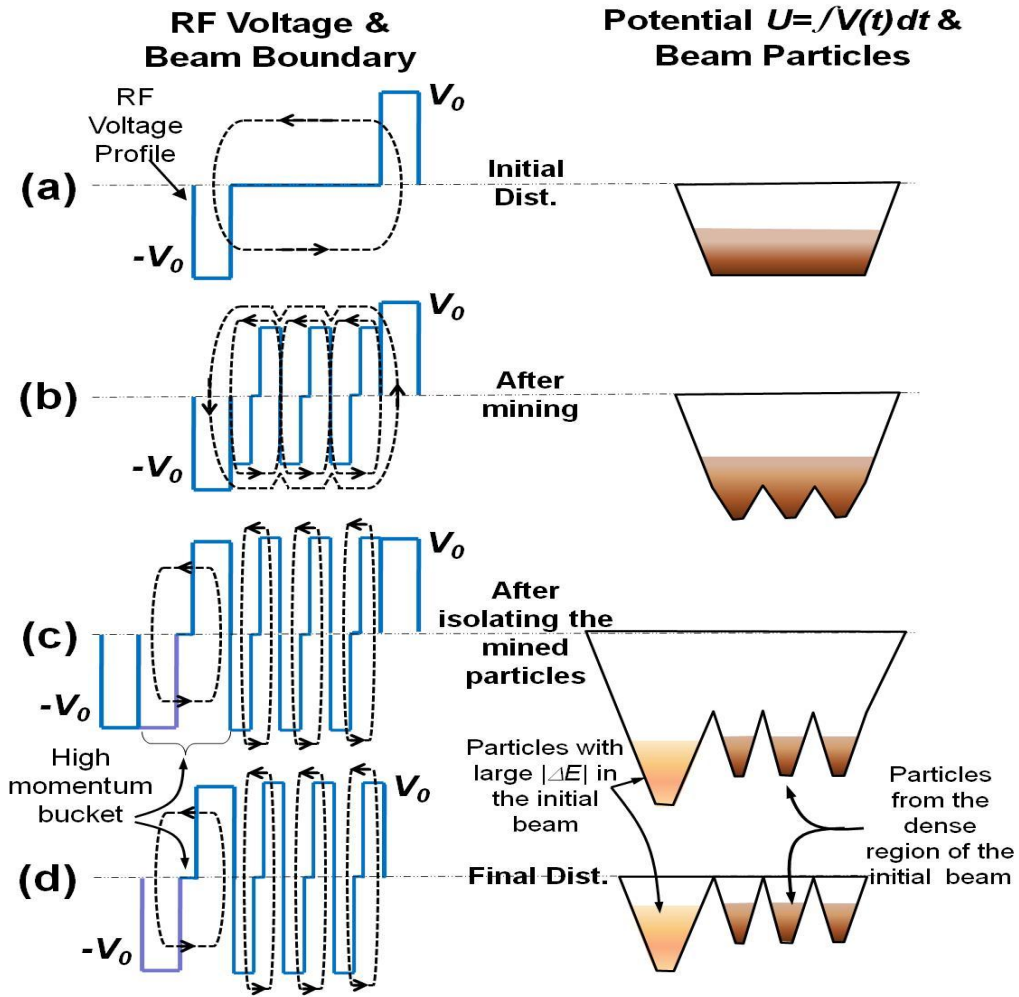

Fig. 11: Schematic illustrating the principle of longitudinal momentum mining for three parcels using barrier buckets. Barrier rf voltage waveforms (solid blue lines) and beam particle boundaries in ($\Delta E, \tau$)-phase space (dashed lines) are shown on the left. The potential $\int V(t) dt$ and the beam particles in it in each case is shown in the cartoon on the right.

The LMM on the beam cooled with e-cooling showed that the mined buckets found to have particles with low longitudinal as well as low transverse momenta[4].

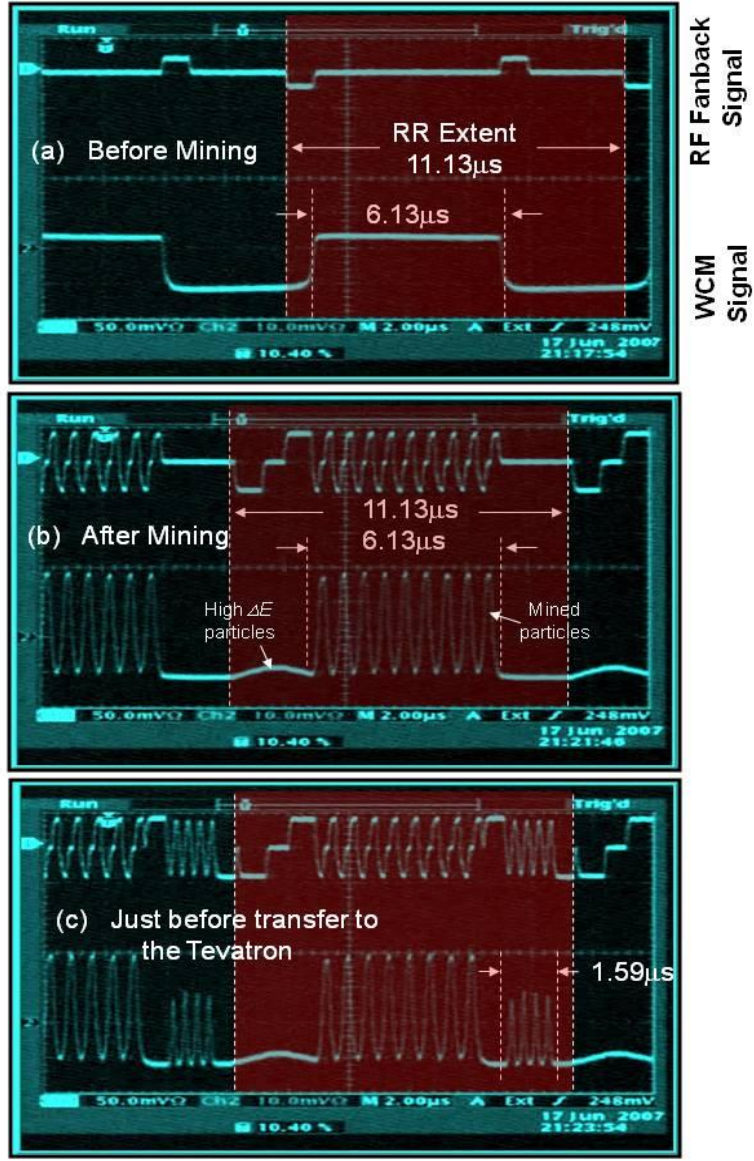

Fig. 12: The experimental data on longitudinal momentum mining in the Recycler with the antiproton beam a) before mining, b) after mining, c) just before the first transfer to the Tevatron. The description of the shaded region and traces are similar to Fig. 8.

---

[4] In the absence of e-cooling this is not true. In any case, the LMM guarantees mining of the particles with low longitudinal momenta.

Figure 12 shows scope pictures for LMM in the RR just before a typical Tevatron proton-antiproton store. The two traces in each case are the rf fan-back and WCM signals. Figure 12(a) shows the initial cold stack of about $3.5\times 10^{12}$ antiprotons captured in a rectangular barrier bucket of $T_2$=6.13 μs, $T_1$=0.903 μs and with rf pulse height of about 1 kV. The measured longitudinal emittance of the stack was about 68 eVs (95%). The mining is carried out with nine mining buckets with a predetermined area. The size of a mining bucket is decided based on the needs of the Tevatron and MI acceptance. The area of the mini-bucket was chosen to be about 8 eVs. The beam after mining and separating particles with high $|\Delta E|$ is shown in Fig. 12(b). Subsequently, the first parcel from the right is further divided into four 2.5 MHz bunches after moving it to the extraction region of the RR. Just a few seconds before extraction to the Tevatron, an anti-bucket is grown adiabatically to leave behind any un-captured antiprotons and keep them away from the extraction region as shown in Fig. 12(c). The anti-bucket is quite important to help sending clean bunches to the Tevatron by eliminating any undesirable DC beam into the MI. During the entire barrier rf manipulations importance is given to preserve the emittance as well as antiproton economy.

A few improvements have been added to the process explained above. For example, we found that it may be more advantageous to eliminate the high momentum bucket and keep a small amount of un-captured antiprotons freely moving in the ring to help stability of the mined beam. But, it is important to note that if the antiproton beam is not sufficiently cold before mining (for example, $\varepsilon_l$ (95%)≥70 eVs), it is highly advisable to follow the conventional LMM explained earlier in view of antiproton economy and good transfer efficiencies.

Another improvement, strip mining, is used whenever the Tevatron demands only a part of the antiproton stack. The stack is divided into two parts prior to the mining and the rest is quite similar to the process explained above.

Figure 13(b) shows the measured antiproton intensity, beam peak density and $\varepsilon_H$ with flying wires in the RR for a typical mining case during a typical beam transfer to the Tevatron. The sudden increase in initial peak density represents the transition from the un-mined state to the mined state (see Fig. 12(a) and Fig 12(b)). Then, the beam is held in the mined state for a minimum of two minutes with e-cool turned on to cool the beam further. The cooling continues until the last transfer is taken out of the RR. A slow increase in the transverse emittance is often observed due to a small transverse kick on the beam from the RR extraction kickers. Measurements on the extracted beam showed that the longitudinal emittance of the 2.5 MHz bunch is about 1.0 eVs and the intensity is ~1/36 of the total extracted beam within a fluctuation of about 2.5%.

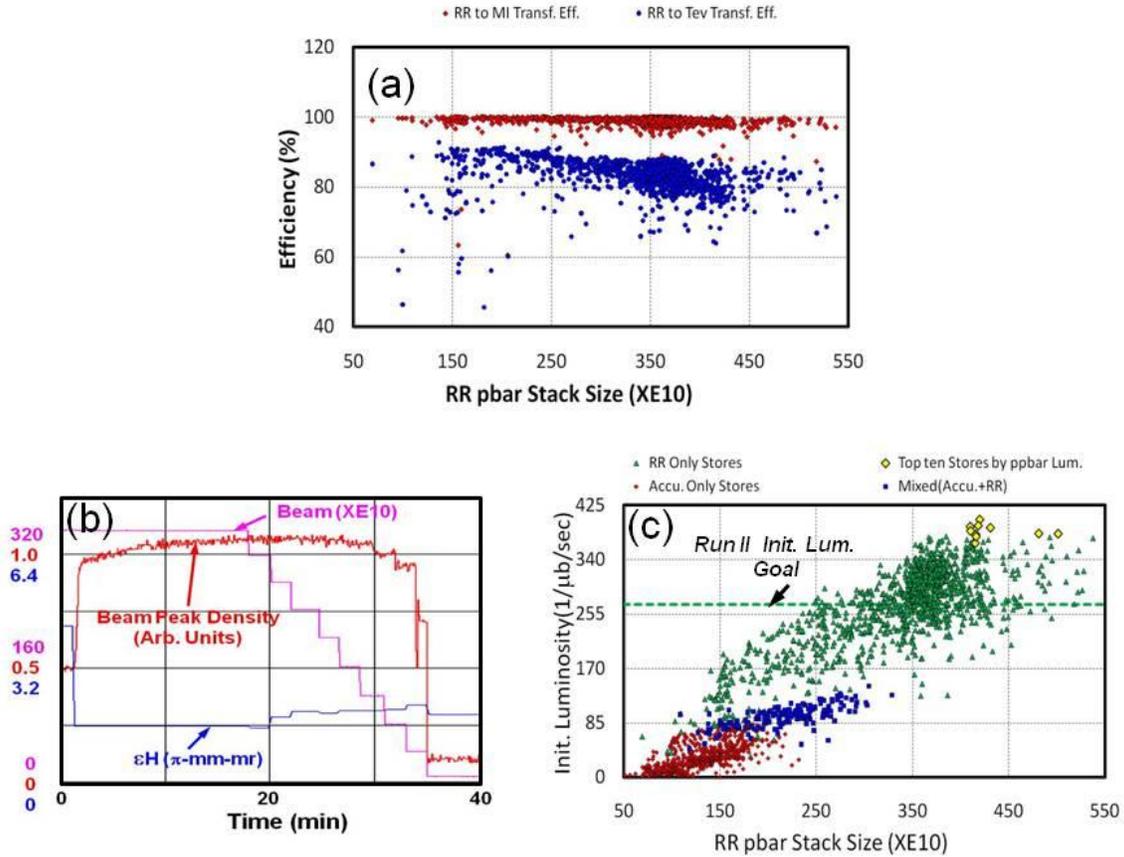

Fig. 13: Measured a) efficiencies, b) antiproton beam intensity, average transverse emittance using flying wires and relative peak density during a typical mining and transfers to the Tevatron, c) initial luminosities as a function of RR antiproton stack sizes for all proton-antiproton stores for Run II. The top ten proton-antiproton stores in the Tevatron by their peak luminosity are indicated by yellow diamonds. The Run II luminosity goal [34] is shown by the dashed line for comparison.

Measured RR to MI and RR to Tevatron beam transfer efficiencies as a function of antiproton stack size for all the proton-antiproton stores since August 2007 are shown in Fig.

13(a) by red and blue points respectively. The average transfer efficiency between the RR and MI was close to 100% while the RR to Tevatron efficiency was about 83% (with the highest efficiency of about 96% from an initial stack size around $1.0\times10^{12}$ antiprotons). A major contribution to this antiproton loss in the MI came from the bunch coalescing efficiency at 150 GeV.

A comparison of initial proton-antiproton luminosities in the Tevatron for Run II is shown in Fig. 13(c). One can see clear distinction between Tevatron proton-antiproton initial luminosities for stores with Accumulator, Accumulator in combined with RR (mixed stores) and RR for same amount of available antiprotons –stores with RR had about a factor of two higher luminosities as compared to those with Accumulator. Data clearly shows that during these years the Tevatron performance exceeded the Run II design luminosity goal of $270\times10^{30}/cm^2/sec$ [34] by about 50%.

*Acknowledgements*

The author would like to thanks a number of colleagues in the Accelertor Division. Special thanks to Dave Wildman, Brian Chase, Cons Gattuso, Jim MacLachlan, Joe Dey, Eddy Nathan, Paul Joireman, Martin Hu and Jim Zagel and many others for fruitful discussions and their inputs. I also thank the Fermilab Accelerator Division Operation group, without their cooperation and help any of the rf manipulation tests could not have been done. My special thanks to Shreyas Bhat for careful reading of this document.

**References**

[1]  Design Report Tevatron 1 Project
     http://lss.fnal.gov/archive/design/fermilab-design-1984-01.pdf

[2]  J.E. Griffin *et al.,* IEEE, Trans. Nucl., Sci., NS-30, 3502 (1983).

[3]  A. G. Ruggiero, CERN 68-22, Intersecting Storage Rings Division, 4 June, 1968 (and references there in); A. G. Ruggiero, ISR-Th/67-68.

[4]  "Applications of barrier bucket RF systems at Fermilab", C. M. Bhat, "Proceedings of RPIA 2006 Recent Progress In Induction Accelerators" 7-10 Mar 2006, KEK, Tsukuba, Japan edited by K. Takayama, (2007) page 45-59; FERMILAB-CONF-06-102-AD

[5]  G. Jackson, Fermilab-TM-1991, November, 1996.
    http://lss.fnal.gov/archive/test-tm/1000/fermilab-tm-1991.pdf

[6]  "Fermilab Main Injector Technical Design Handbook" (Internal Report, 1994).

[7]  "Theory of RF acceleration and RF noise" G. Dome, CERN SPS/84-13(ARF) (1984) (and references there in).

[8]  S. Y. Lee, *Accelerator Physics*, **1**st edition (World Scientific, Singapore, 1999), Chap. V, p.305.

[9]  S.Y. Lee and K. Y. Ng, Phys. Rev. E, Vol. 55, 5992 (1997).

[10] "Longitudinal Stability of Recycler bunches Part I: Threshold of loss of Landau Damping", T. Sen, C. M. Bhat and J.-F. Ostiguy, FERMILAB-TM-2431-APC, June 9, 2009.

[11] J. A. MacLachlan, HEACC'98, XVII International Conference on High Energy Accelerators, Dubna, 184, (1998); J. A. MacLachlan, Fermilab Report No FN-529, (1989) (Unpublished); (private communications 2004); The latest version of the code is available at http://www-ap.fnal.gov/ESME/.

[12] "Current DSP Applications in Accelerator Instrumentation and RF", B. Chase, A. Mason, and K. Meisner, ICALEPS'97 (1997) (unpublished)
    http://www.aps.anl.gov/News/Conferences/1997/icalepcs/paper97/p230.pdf;
    P. Joireman, B. Chase and K. Meisner, since 2004 a significant improvement is made in RRLLRF (private communications).

[13] J. P. Marriner (2001-2002, unpublished).

[14] J. E. Dey and D. W. Wildman, PAC1999, p 869; (also private commutations with Dave Wildman and Joe Dey).

[15] "A first look at the longitudinal emittance in the Fermilab Recycler Ring", C. M. Bhat, MI Note 287, August 2002; (Beams-Doc-376-v1).

[16] J. P. Marriner and C. M. Bhat (unpublished, 2002).

[17] J. Dey, S. Dris, T. Kubick and J. Reid, PAC2003, p1204; (and private communications with J. Dey, 2011).

[18] C.M. Bhat and K. Y. Ng, FERMILAB-CONF-03-395-T (Oct 2003) (to be published in Proc. of Factories 2003, Stanford, California, October, 2003).

[19] J. Crisp, H. Hu and K.Y. Ng, HB2006, p 244.

[20] M. Hu et. al., PAC07, p458; "Adaptive RF Corrections for Recycler" N. Eddy et. al., a report in "All Experimenters Meetings Special Reports," February 26, 2007 (unpublished) http://www.fnal.gov/directorate/program_planning/all_experimenters_meetings/special_reports/

[21] D. Boussard, CERN SPS/86-11(ARF), Geneva 1986.

[22] V. Balbekov and S. Nagaitsev, EPAC2004, p 791.

[23] C. M. Bhat and J. Marriner, PAC2003, p 514.

[24] D. R. Broemmelsiek, et. al., EPAC04, p 794; D. R. Broemmelsiek (Private communications).

[25] "Beam Longitudinal Tomography for a Barrier Bucket" A. Burov, Beams-doc-2866, Aug. 2007; "Reconstruction of Longitudinal Phase-space by Monte Carlo Method" C. M. Bhat, Beams-doc-2884, Oct. 2007.

[26] MeiQin Xiao, "A console application program to measure longitudinal emittance of the antiprotons in 2.5 MHz buckets" (unpublished).

[27] B. Chase, (private communications, 2002)

[28] C. M. Bhat, PAC2003, p 2345; C. M. Bhat, PAC2005, p 1093; 'Pbar stacking in the Recycler: Longitudinal Phase-Space Coating", C. M. Bhat, Beams-Doc-2057-V1, December 2005; J. MacLachlan, FERMILAB-Conf-00/117, June 2000; C. M. Bhat, EPAC2008, p 307; "Stacking of antiprotons by morph-merging," P. Joireman and B. Chase (private communications 2007).

[29] S. Nagaitsev et.al., Phys. Rev. Lett. 96, (2006) p 044801.

[30] S. M. Pruss, "RR the orbit compensation to the Main Injector cycles" 2002-2004, (private communications); P.Lebrun, "Recylcer Orbit Length Optimization, June 29 Status report" June 2005 (Beams-Doc-1886-v1).

[31] N. Eddy and J. Crisp, FERMILAB-Conf-06-100-AD; N. Eddy, J. L. Crisp and M. Hu, AIP Conf. Proc. 868, p293 (2006); N. Eddy, J. L. Crisp and M. Hu, EPAC'08 (2008) p3254.

[32] C. M. Bhat et. al. PAC2005, p 1763.

[33] C. M. Bhat, Phys. Letts. A 330 (2004) 481-486; C. M. Bhat, EPAC2004, p 236; C. M. Bhat, FERMILAB-FN-746, 2004, http://lss.fnal.gov/archive/test-fn/0000/fermilab-fn-0746.pdf

[34] P.C. Bhat and W. J. Spalding, *Proc. 15th Topical Conf. on Hadron Collider Physics*, East Lansing, Michigan, 2004, *AIP Conf.Proc.*753:30 (2005).